\documentclass[review]{elsarticle}
\usepackage{graphicx} 
\usepackage{bmpsize}
\usepackage{dcolumn}% Align table columns on decimal point
\usepackage{bm}% bold math
\usepackage{hyperref}% add hypertext capabilities
\usepackage{braket}
\usepackage{xcolor}
\usepackage{mathrsfs}
\usepackage{setspace}
\usepackage{colortbl,booktabs}
\usepackage{lineno,hyperref}
\usepackage{geometry}
\modulolinenumbers[5]
\journal{Communications in Nonlinear Science and Numerical Simulation}

\biboptions{sort}
\begin{document}

\title{Fast influencers in complex networks}

\author{Fang Zhou}
\address{Institute of Fundamental and Frontier Sciences, University of Electronic Science and Technology of China, Chengdu 610054, PR China}

\author{Linyuan L\"u}%
\ead{linyuan.lv@uestc.edu.cn}
\address{Institute of Fundamental and Frontier Sciences, University of Electronic Science and Technology of China, Chengdu 610054, PR China}
\address{Alibaba Research Center for Complexity Sciences, Hangzhou Normal University, Hangzhou 311121, China}

\author{Manuel Sebastian Mariani}
\ead{manuel.mariani@business.uzh.ch}
\address{Institute of Fundamental and Frontier Sciences, University of Electronic Science and Technology of China, Chengdu 610054, PR China}
\address{URPP Social Networks, Universit\"at Z\"urich, Switzerland}

%\date{}% It is always \today, today,

\begin{abstract}
Influential nodes in complex networks are typically defined as those nodes that maximize the asymptotic reach of a spreading process of interest. However, for practical applications such as viral marketing and online information spreading, one is often interested in maximizing the reach of the process in a short amount of time. The traditional definition of influencers in network-related studies from diverse research fields narrows down the focus to the late-time state of the spreading processes, leaving the following question unsolved: which nodes are able to initiate large-scale spreading processes, in a limited amount of time? Here, we find that there is a fundamental difference between the nodes -- which we call ``fast influencers" -- that initiate the largest-reach processes in a short amount of time, and the traditional, ``late-time" influencers. 
Stimulated by this observation, we provide an extensive benchmarking of centrality metrics with respect to their ability to identify both the fast and late-time influencers. We find that local network properties can be used to uncover the fast influencers. In particular, a parsimonious, local centrality metric (which we call social capital) achieves optimal or nearly-optimal performance in the fast influencer identification for all the analyzed empirical networks. Local metrics tend to be also competitive in the traditional, late-time influencer identification task.
\end{abstract}

\maketitle

%\tableofcontents

\section{Introduction}

Diffusion processes on networks can describe phenomena as diverse as the spreading of infectious diseases~\cite{pastor2015epidemic}, the life cycle of products and services~\cite{peres2010innovation}, the spreading of  behaviors~\cite{centola2010spread}, among others.
A long-standing idea in social sciences and marketing research is that in a social network, a minority of individuals have the possibility to disproportionately impact on the diffusion of an innovation~\cite{katz1955personal,watts2007influentials}.
While marketing researchers have investigated various categories and traits of such influential individuals -- including domain expertise~\cite{king1970overlap,tanase2018identification}, wide market information~\cite{nejad2014influentials}, and opinion leadership~\cite{katz1955personal} -- the physics and network science literature have mostly focused on the individuals' ability to reach a large number of individuals by benefiting from their favorable position in the network of contacts~\cite{lu2016vital}.

In this context, the ``influencers"~\cite{pei2018theories} (or ``influentials" ~\cite{watts2007influentials}, or ``influential spreaders"~\cite{kitsak2010identification}) are typically defined as those nodes who, once they initiate a spreading process on a network of contacts, can reach and ``infect" a large portion of the system. Following the seminal works by Kempe et al.~\cite{kempe2003maximizing} and Kitsak et al.~\cite{kitsak2010identification}, a large number of papers have attempted to identify the influencers based on network structural analysis and centrality metrics~\cite{kitsak2010identification,borge2012absence,morone2015influence,lu2016h,lu2016vital,radicchi2016leveraging,radicchi2017fundamental,pastor2017topological,iannelli2018influencers} (see~\cite{lu2016vital} for a review). In this stream of works, the ground-truth influence of a given ``seed" node (or equivalently, its spreading ability) is defined as the average number of nodes that are eventually reached by independent diffusion processes initiated by that seed node~\cite{lu2016vital}. In the following, we refer to this property as node \emph{late-time influence}, and we refer to nodes with large late-time influence as \emph{late-time influencers}.

Based on this framework, scholars evaluated the performance of network centrality metrics according to their ability to reproduce the ranking of the nodes by their late-time influence.
The results have fostered the debate~\cite{lu2016vital,pastor2017topological,hu2018local} on whether local centrality metrics~\cite{chen2012identifying,chen2014path,morone2015influence,liu2016identify,lu2016h,hu2018local} (i.e., metrics that infer nodes' centrality based purely on local information) or global metrics~\cite{kitsak2010identification,de2014role,ren2014iterative,radicchi2016leveraging,iannelli2018influencers} (i.e., centrality metrics that take into account the whole network structure) should be preferred for the influencer identification. Local metrics are faster to be computed and can provide optimal or nearly optimal performance~\cite{lu2016h,lu2016vital,hu2018local}, yet more computationally-expensive global metrics might provide a more accurate quantification of node influence at the critical point~\cite{radicchi2016leveraging} and above~\cite{iannelli2018influencers}.

However, the definition of node ground-truth influence adopted so far implicitly narrows down the problem's focus to the stationary state of the dynamics.
On the other hand, for marketing applications, it is vital for companies to accumulate customers or adopters as rapidly as possible~\cite{helm2000viral}. 
Besides, the early stage of spreading processes plays an essential role in the dissemination of low-credibility content in online social platforms~\cite{shao2018spread}; therefore, to design effective prevention strategies, it becomes critical to gauge the nodes' ability to disseminate information in a small amount of time.
Previous studies have addressed the impact of network topology and temporal effects on diffusion speed~\cite{karsai2011small,scholtes2014causality}, the design of local strategies specifically aimed at fast spreading~\cite{zhou2006behaviors}, and the identification of nodes that tend to be infected earlier than the others~\cite{christakis2010social}.
However, they did not address the fundamental question whether strategies that are effective in maximizing the late-time impact of a process are also effective at early times, and they did not investigate whether well-performing centrality metrics are only competitive at specific timescales.  

To fill these gaps, we introduce here the \emph{time-dependent influence} of a given ``seed" node as the average number of nodes that have been infected during a finite number $t$ of diffusion steps, by independent diffusion processes initiated by that seed node. We refer to the time-dependent influence after few steps of the dynamics as the \emph{early-time influence} of the nodes, and to the nodes with large early-time influence as \emph{fast influencers}.

We find that the nodes' late-time and early-time influence are positively correlated, yet this correlation is far from perfect. The noisy relation between the two types of influence reveals that strategies that aim at maximizing the late-time impact of a spreading process (i.e., those typically considered in the literature) can be suboptimal for maximizing its early-time impact. The late-time influencers may indeed initiate large-scale processes that take a long time to diffuse, whereas nodes with a lower late-time influence might initiate faster processes.

A previously unexplored question, therefore, emerges: how to best identify the fast influencers?
We find that local network information can be used to accurately quantify the early-time influence of the nodes.
In particular, a simple combination of node degree and neighbors' degrees is optimal or nearly optimal in quantifying the early-time influence of the nodes. Our extensive comparative analysis reveals that other local metrics (including LocalRank~\cite{chen2012identifying} and the Dynamics Sensitive Centrality~\cite{liu2016locating}) can be similarly competitive at early times. On the other hand, global metrics (such as the $k$-core centrality~\cite{kitsak2010identification} and the closeness centrality~\cite{freeman1978centrality}) can be only competitive in the late-time influencer identification, yet for some datasets and dynamics parameter values, local metrics outperform global metrics also in the late-time influencer identification.

Our findings shed light on the fundamental difference between nodes' early-time and late-time influence.  Importantly, they suggest that compared to global metrics, local metrics for the influential nodes identification might be competitive for a broader range of dynamical processes.

The rest of the paper is organized as follows: Section~\ref{sec:theory} introduces the traditional late-time influencer identification problem together with relevant centrality metrics and the datasets analyzed here;
Section~\ref{sec:difference} presents the fundamental difference between fast and late-time influencers; Section~\ref{sec:benchmarking} presents the results of an extensive benchmarking of centrality metrics with respect to their ability to single out fast and influential spreaders. Finally, Section~\ref{sec:conclusions} summarizes our main results and outlines open directions.
                                                                        
\section{Theory and methods}
\label{sec:theory}

The goal of this Section is to introduce the traditional definition of influencers based on epidemic spreading processes (Section~\ref{sec:influencer}), the centrality metrics that have been studied to identify the influencers (Section~\ref{sec:centrality}), and the empirical networks analyzed in this paper (Section~\ref{sec:empirical}).

\subsection{The traditional definition of influencers based on spreading models}
\label{sec:influencer}

The problem of identifying influential nodes (hereafter, ``influencers"~\cite{pei2018theories}) in a network generally refers to the problem of identifying a small set of nodes, $\mathcal{S} \subset \mathcal{V}$ (where $\mathcal{V}$ denotes the set of all nodes, and $|\mathcal{S}|\ll |\mathcal{V}|$), that maximizes a certain target function, $f(\mathcal{S})$. Such a function could be a function of the topology of the network (``structural influence maximization"), or a function that depends on the interplay between network topology and a given dynamical process on the network (``functional influence maximization")~\cite{lu2016vital}. In this work, we consider the functional influence maximization problem, where the influencers are traditionally defined as the nodes that maximize the late-time reach of the dynamic process in exam.

Given a network, the late-time influence of the nodes depends on the target dynamics under consideration. 
In line with a popular stream of physics literature~\cite{kitsak2010identification,lu2016h,lu2016vital,liu2016locating,radicchi2016leveraging,iannelli2018influencers,hu2018local}, we focus here on the Susceptible-Infected-Recovered dynamics (SIR).
The SIR dynamics on a given network starts from an initial configuration where some of the nodes (typically referred to as ``seed nodes") are in the infected (I) state. At each time step, each infected node can infect each of its susceptible (S) neighbors with probability $\beta$. In addition, each infected node can recover with probability $\mu$. The dynamics ends when all the nodes are in the recovered (R) state. 
For convenience, in the following, we fix $\mu=1$, and express the results in terms of the ratio $\lambda/\lambda_c$ between the spreading rate $\lambda=\beta/\mu$ and its critical value $\lambda_c$. In line with previous studies~\cite{lu2016vital}, to compute $\lambda_c$, we use the analytic result obtained with a degree-based mean-field approximation~\cite{pastor2015epidemic}: $\lambda_c=\braket{k}/(\braket{k^2}-\braket{k})$, where $\braket{k}$ and $\braket{k^2}$ denote the average degree and the average squared degree, respectively.

Importantly, only a portion of the nodes end up in the recovered state. We focus here on processes with one single seed node at time zero; clearly, the number of nodes in the recovered state depends on the chosen seed node. For each node $i$, its late-time influence $q_i(\infty)$ is defined as the average \emph{asymptotic} fraction of nodes in the recovered state for processes initiated with node $i$ as the seed node.
The influential spreaders (here, late-time influencers) are defined as the nodes with the largest influence $q(\infty)$. Accordingly, network centrality metrics are typically compared with respect to their ability to single out the high-$q(\infty)$ nodes~\cite{lu2016vital}.

\subsection{Centrality metrics}
\label{sec:centrality}

Centrality metrics generally aim to identify the most ``important" nodes in a network~\cite{liao2017ranking}.
Their performance in the influencer identification task is traditionally evaluated according to their ability to identify the late-time influencers defined in the previous Section~\cite{lu2016vital}.
We describe here the metrics analyzed in this paper. We denote as $\mathsf{A}$ the adjacency matrix of the undirected network: $A_{ij}=1$ if a link between node $i$ and $j$ exists.

\paragraph{Degree}
The degree of a node is defined as its total number of neighbors. In terms of the adjacency matrix, we have~\cite{newman2018networks}
\begin{equation}
k_i = \sum_{j=1}^N A_{ij}.
\end{equation}

\paragraph{H-index}

The $H$-index centrality was recently validated as an effective metric for the influencer identification~\cite{lu2016h}.
To define the $H$-index, let us denote by $k_{i,1}$,$k_{i,2}$,$k_{i,3}$,...,$k_{i,k_i}$ the degrees of the neighbors of a given node $i$. Then, the $H$-index of node $i$ is $h_i$ if and only if $h_i$ is the largest integer number such $h_i$ neighbors of node $i$ have degree larger than or equal to $h_i$. The construction of the $H$-index can be generalized to construct higher-order centrality metrics, which leads to an iterative procedure to construct the $k$-core centrality (see below) starting from degree~\cite{lu2016h}.

\paragraph{LocalRank}
The LocalRank score $L_i$ of node $i$ is defined as~\cite{chen2012identifying}
\begin{equation}
L_i = \sum_{j=1}^N A_{ji}\sum_{l=1}^N A_{lj}\,N_l,	
\end{equation}
where $N_l$ is the number of the nearest and the next nearest neighbors of node $l$. In practice, the LocalRank score takes into account local information up to distance four from the focal node.

%\paragraph{Collective Influence}

\paragraph{Dynamics Sensitive Centrality}
The Dynamics Sensitive Centrality (DSC)~\cite{liu2016locating} aims to estimate analytically the time-dependent influence of the nodes. Denoting as $s_i(t)$ the estimated influence of node $i$ at time $t$, by analytically estimating the fraction of newly infected nodes at each time step, one finds analytically that~\cite{liu2016locating}
\begin{equation}
\mathbf{s}(t)=(\beta\,\mathsf{A}+ \beta\,\mathsf{A}\,\mathsf{H}+\dots+\beta\,\mathsf{A}\,\mathsf{H}^t)^T\,\mathbf{e},
\end{equation}
where $\mathsf{H}=\beta\,\mathsf{A}+(1-\mu)\,\mathsf{I}$, $\mathsf{I}$ denotes the $N\times N$ identity matrix, and $\mathbf{e}$ denotes the vector whose $N$ components are all equal to one. The metric presents the disadvantage that in order to be applied, one needs to know the values of $\beta$ and $\mu$, which is typically not the case in real applications. However, we found that its performance is not strongly dependent on the value of $\beta$; for this reason, we set $\mu=1$ and   $\beta=\lambda_c$.

\paragraph{$k$-core}

In an undirected unweighted graph, the $k$-core score of the nodes is defined in terms of a network pruning process~\cite{kitsak2010identification}. One starts by removing all nodes with a degree equal to one. After this step is completed, some nodes that had $k>1$ may end up with a degree equal to one: these nodes are removed as well. One keeps removing the nodes that end up with $k=1$ iteratively until no node with $k=1$ is left in the network. All the nodes removed throughout this process are assigned a $k$-core score equal to one.
One performs an analogous procedure for the nodes with $k=2, 3, \dots$, thereby identifying all the nodes with a $k$-core score equal to $2, 3, \dots$. The calculation ends when all the nodes have been assigned a $k$-core score.

\paragraph{Eigenvector centrality} 
The vector of eigenvector-centrality scores is defined as the eigenvector of the adjacency matrix $\mathsf{A}$ associated with $\mathsf{A}$'s largest eigenvalue $\lambda_{max}$~\cite{Phillip2001eigenvector}. In formulas,
\begin{equation}
 \mathsf{A}\, \mathbf{x} =\lambda_{max}\,\mathbf{x}   
\end{equation}
where $\lambda_{max}$ is the biggest eigenvalue of matrix A, and x is the corresponding eigenvector.

\paragraph{Betweenness}
The betweenness centrality score $B_i$ of node $i$ is defined as the average ratio between the number $n_{st}^i$ of shortest paths between two nodes $s$ and $t$ that pass through a focal node $i$, and the total number $g_{st}$ of shortest paths between node $s$ and $t$~\cite{Brandes2001betweenness}. In formulas,
\begin{equation}
B_i = \sum_{s\neq i \neq t} \frac{n_{st}^i}{g_{st}}.
\end{equation}

\paragraph{Closeness}

In a connected network, the closeness centrality score~\cite{freeman1978centrality} of node $i$ is defined as the reciprocal of its average distance from the other nodes in the network, where the distance is determined by the shortest-path length. In formulas,
\begin{equation}
d_i= \frac{1}{N-1}\sum_{j=1}^{N} d_{ij},
\end{equation}
where $d_{ij}$ is the length of the shortest paths that connect node $i$ and $j$;
the closeness centrality score of node $i$ is defined as
\begin{equation}
    C_i ={d_i}^{-1}.
\end{equation}
If the network is disconnected, one first computes the average distance $d_i$ of node $i$ from the nodes that belong to the same network component $\mathcal{G}_i \subset \mathcal{G}$. In formulas,
\begin{equation}
d_i= \frac{1}{|\mathcal{G}_i|-1}\sum_{j \in \mathcal{G}_i} d_{ij},
\end{equation}
where $|\mathcal{G}_i|$ denotes the number of nodes that belong to $\mathcal{G}_i$. Then, one computes the closeness centrality score of node $i$ as~\cite{freeman1978centrality}
\begin{equation}
    C_i=\frac{|\mathcal{G}_i|-1}{(N-1)}.d_{i}^{-1}.
\end{equation}

\begin{table}[t]
\resizebox{\textwidth}{!}{%
\begin{tabular}{|c|c|c|c|c|c|c|}

\hline
\textbf{Dataset name} & $N$ & $L$ & $\braket{k}$  & $\braket{k^2}$  &$\lambda_c$  & URL \\
\hline
\textbf{Amazon} & 334863 & 925872& 5.5 & 63.7&0.094 &\url{http://konect.uni-koblenz.de/networks/com-amazon}\\
\hline
\textbf{Cond-mat} & 27519&116181  & 8.4& 188.1  &0.047 &\url{http://networkscience.cn/index.php/data/}\\
\hline
\textbf{Email-Enron} &36692 & 183831 & 10.0 &1402 &0.0071 &\url{http://networkscience.cn/index.php/data/}\\
\hline
\textbf{Facebook} & 63731 & 817035 & 25.6 &2256.5 &0.011 &\url{http://networkscience.cn/index.php/data/}\\
\hline
arenas-email & 1133 & 5451 & 9.6 & 179.8 & 0.0564 &\url{http://konect.uni-koblenz.de/networks/arenas-email} \\
\hline
arenas-pgp & 10680 & 24316 & 4.5 & 85.9 & 0.0553 &\url{http://konect.uni-koblenz.de/networks/arenas-pgp}\\
\hline
Route views & 6474 & 12572 & 3.9 & 640 &0.006 &\url{http://konect.uni-koblenz.de/networks/as20000102}\\
\hline
ca-AstroPh & 18771 & 198050 & 21.1 & 1379 & 0.015 &\url{http://konect.uni-koblenz.de/networks/ca-AstroPh}\\
\hline
Human protein  & 3023 & 6149 & 4.06 & 62.8&0.069 &\url{http://konect.uni-koblenz.de/networks/maayan-vidal}\\
\hline
petster-hamster & 2426 & 16631 & 13.7 & 582.9 &0.024 &\url{http://konect.uni-koblenz.de/networks/petster-hamster}\\
\hline
 reactome& 6229&146160&43.71& 6708.3 &0.007&\url{http://konect.uni-koblenz.de/networks/reactome}\\
\hline
Stelzl& 1615&3106& 3.8&65.6&0.061 &\url{http://networkrepository.com/maayan-Stelzl.php}\\
\hline
vidal&2783& 6007 & 4.3&68.1&0.067&\\
\hline
Yeast& 2375& 11693&9.8&336.9&0.030 &\url{http://www.linkprediction.org/index.php/link/resource/data}\\
\hline
\end{tabular}}
\caption{Details of the analyzed empirical networks: number of nodes $N$; number of links $L$; average degree $\braket{k}$; average square degree $\braket{k^2}$; critical value $\lambda_c$ of the SIR model parameters' ratio $\lambda=\beta/\mu$; URL where the dataset can be downloaded. }
\label{tab}
\end{table} 

\subsection{Empirical datasets}
\label{sec:empirical}

We analyze here the same four networks that were analyzed in the review article~\cite{lu2016vital}, plus 10 smaller networks. We denote the four largest networks as \emph{Amazon, Cond-mat, Email-Enron, Facebook}. \emph{Amazon}~\cite{amazon} is the co-purchase network from the Amazon website. Its nodes represent products and the links between two nodes indicate that the two products have been frequently bought together. \emph{Cond-mat}~\cite{condmat} is a collaboration network of scientists. Its nodes represent scientists and a link between two nodes means that the two scientists have co-authorized at least one paper. \emph{Email-Enron}~\cite{email} is a communication network. Its nodes represent email address; there exists a link between two addresses if the two addresses have exchanged at least one email.  \emph{Facebook}~\cite{facebook} is a friendship network from the Facebook social network. Its nodes represent users, and a link between two nodes means that the two users are friends. 
We refer to Table~\ref{tab} for all the details about these four networks and the smaller networks.

\section{The fundamental difference between fast and late-time influencers} 
\label{sec:difference}

In Section~\ref{sec:influencer}, we have seen that the traditional definition of influencers relies on the asymptotic state of the spreading process in exam.
To uncover the fundamental difference between fast and late-time influencers, we introduce the more general problem of finding the optimal influential spreaders at a given finite time $t$. We define the time-dependent influence $q_i(t)$ of node $i$ at time $t$ as the average fraction of nodes in the recovered state for processes initiated by node $i$, after $t$ steps of the diffusion dynamics.
Formally, we recover the traditional definition of (late-time) node influence~\cite{lu2016vital} for\footnote{As the SIR dynamics halts when no infected nodes are left, in numerical simulations, it is sufficient to wait a finite, system-dependent and parameter-dependent computational time in order to measure $q(\infty)$.} $t\to\infty$: $q(\infty):=\lim_{t\to\infty} q(t)$.

\begin{figure*}[t]
\centering
\includegraphics[scale=0.04]{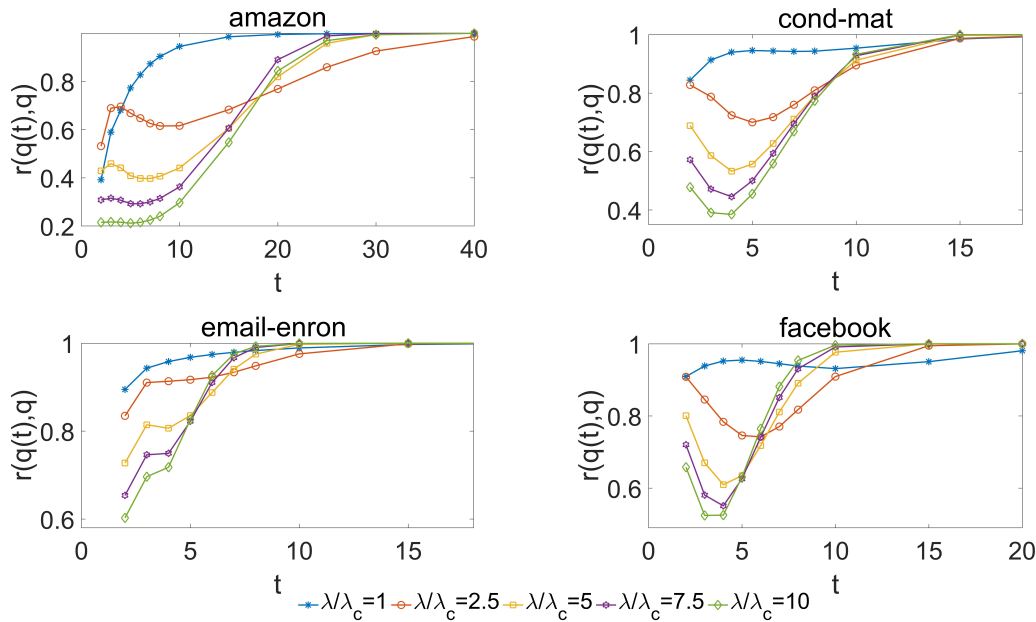}            
\caption{A comparison between node late-time influence $q(\infty)$ and node time-dependent influence $q(t)$, for the four largest empirical networks considered here and different values of the dynamics probability $\lambda$. While positively correlated, the Pearson's linear correlation $r(q(\infty),q(t))$ between $q(\infty)$ and $q(t)$ is substantially smaller than one at early times (intuitively, $t<10$). The discrepancy between $q(\infty)$ and $q(t)$ tends to increase as $\lambda$ increases. }\label{fig:corr} 
\end{figure*}

\begin{figure*}[t]
\centering
\includegraphics[scale=0.125]{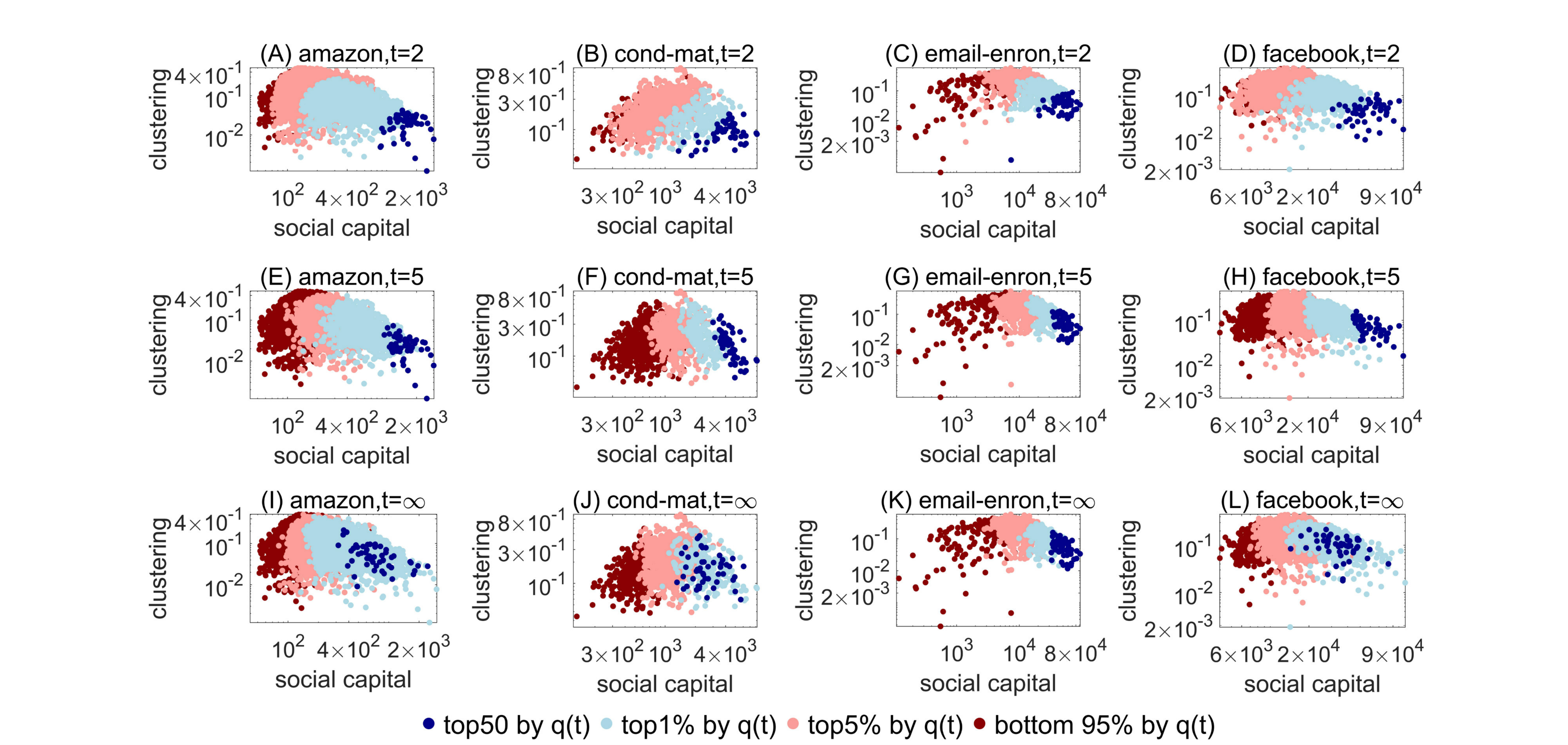}  
\caption{The impact of hubs' social capital on their early-term influence, for the four main datasets analyzed here. The nodes are colored according to their ranking by influence; only the top-$5\%$ nodes by degree $k$ are displayed. Differently from $t=2$ (Panels A--D) and $t=\infty$ (Panels J--L), at $t=5$ (Panels E--H), the most influencer hubs are characterized by a larger social capital $s$: the dark-blue dots (top-$50$ spreaders) tend to accumulate to the right of the panels. By contrast, local clustering does not have a significant impact on node influence.
}
\label{fig:hubs}
\end{figure*}

\subsection{Correlation between early-time and late-time node influence}

Fig.~\ref{fig:corr} shows that while $q(\infty)$ and $q(t)$ are positively correlated, the Pearson's linear correlation $r(q(\infty),q(t))$ between $q(\infty)$ and $q(t)$ is substantially smaller than one at early times (intuitively, $t<10$). The discrepancy between $q$ and $q(t)$ increases as $\lambda$ increases: smaller values of $\beta$ correspond indeed to processes that take less time to reach the final state where there is no infected node left, which results in a better agreement between the early-time and the late-time influence of the nodes. For example, at time $t=5$, in \emph{Amazon}, the correlation $r(q(5),q(\infty))$ is still relatively large for $\lambda=\lambda_c$ [$r(q(5),q(\infty))=0.77$], but it drops to $0.21$ for $\lambda=10\,\lambda_c$. Generally, the larger $\lambda$, the weaker the correlation between node late-time and early-time influence.

\begin{figure*}[t]
\centering
\includegraphics[scale=0.04]{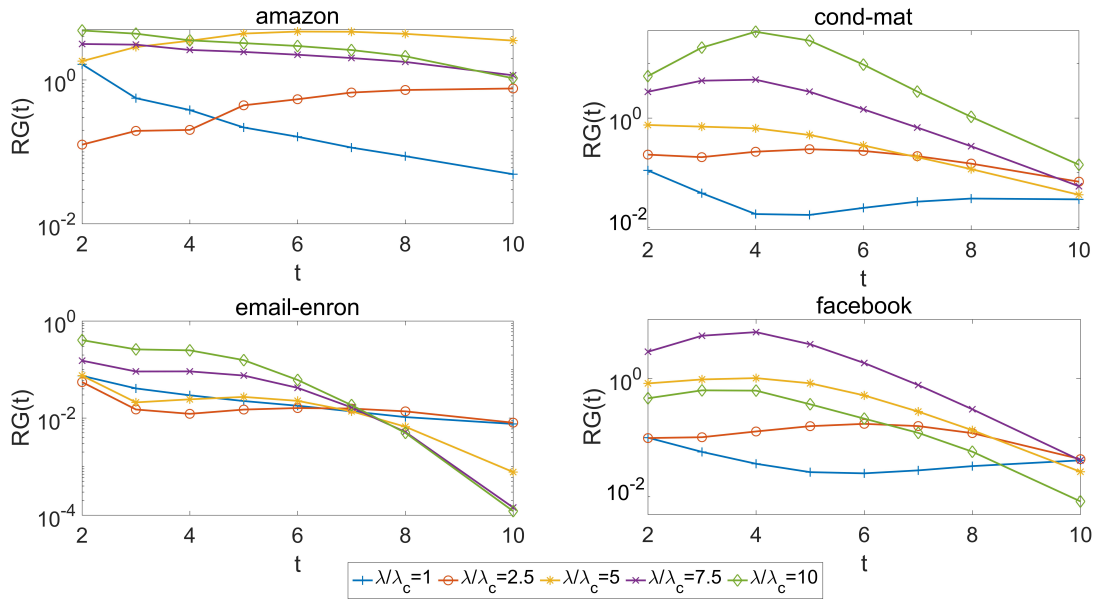}
\caption{Relative gain $RG(t)$ from targeting the top-$0.5\%$ nodes by $q(t)$ as compared to targeting the top-$0.5\%$ nodes by $q(\infty)$, for the first $10$ steps of the dynamics and the four largest networks analyzed here. By definition, the relative gain is always larger than or equal to zero ($RG(t)\to 0$ for $t\to \infty$). Importantly, at early times, we observe a relative gain that is substantially larger than one, which demonstrates that the processes initiated by the top-nodes by $q(t)$ attain a substantially larger reach (improvement by $100\%$), at time $t$, than the processes initiated by the top-nodes by $q(\infty)$.}
\label{fig:gain}
\end{figure*}

Importantly, at a given finite time $t$, targeting the fast influencers (i.e., the nodes with the largest $q(t)$ values) generally leads to a substantial gain in the reach of the process as compared to targeting the traditional, late-time influencers (i.e., the nodes with the largest $q(\infty)$ values). To illustrate this point, we define the relative gain $RG(t)$ from targeting the time-dependent optimal influencers at time $t$ as the difference between the average time-dependent influence $Q(t|t)$ of the top-$0.5\%$ nodes by $q(t)$ and the average time-dependent influence $Q(t|\infty)$ of the top $0.5\,\%$ nodes by $q(\infty)$, normalized by $Q(t|\infty)$. In formulas,
$RG(t):=(Q(t|t)-Q(t|\infty))/Q(t|\infty)$.
Essentially, $RG(t)$ quantifies how larger, at time $t$, are the outbreaks originated by the optimal influencers for time $t$ as compared to the optimal late-time influencers.
The relative gain is, by definition, larger than or equal to zero, and $RG(t)\to 0$ as $t\to\infty$. The observed values of $RG(t)$ (see Fig.~\ref{fig:gain}) indicate that the gains from targeting the optimal influencers for a given time are not marginal; for instance, at time $t=5$ and for $\lambda=5\,\lambda_c$, we observed values of $RG(t)$ as large as $4.41$, $0.49$, $0.03$, and $0.83$ in Amazon, Cond-mat, Email-Enron, and Facebook, respectively. 

Taken together, the results above demonstrate that when seeking to maximize the reach of a spreading process, one faces a trade-off: maximizing the long-term influence of the process (i.e., choosing the seed node with the largest $q(\infty)$) does not provide the optimal influence maximization at a finite time.
This property
sets also time-dependent bounds to the performance of influencer identification metrics~\cite{lu2016vital}: an ideal centrality metric that correlates perfectly with node late-time influence -- i.e., a metric with optimal performance according to the standard benchmarking of influencer identification methods~\cite{lu2016vital} -- is only weakly correlated with node influence at early times.

\subsection{Not all the hubs are fast influencers: the role of neighborhood structure}

\begin{figure*}[t]
\includegraphics[scale=0.26]{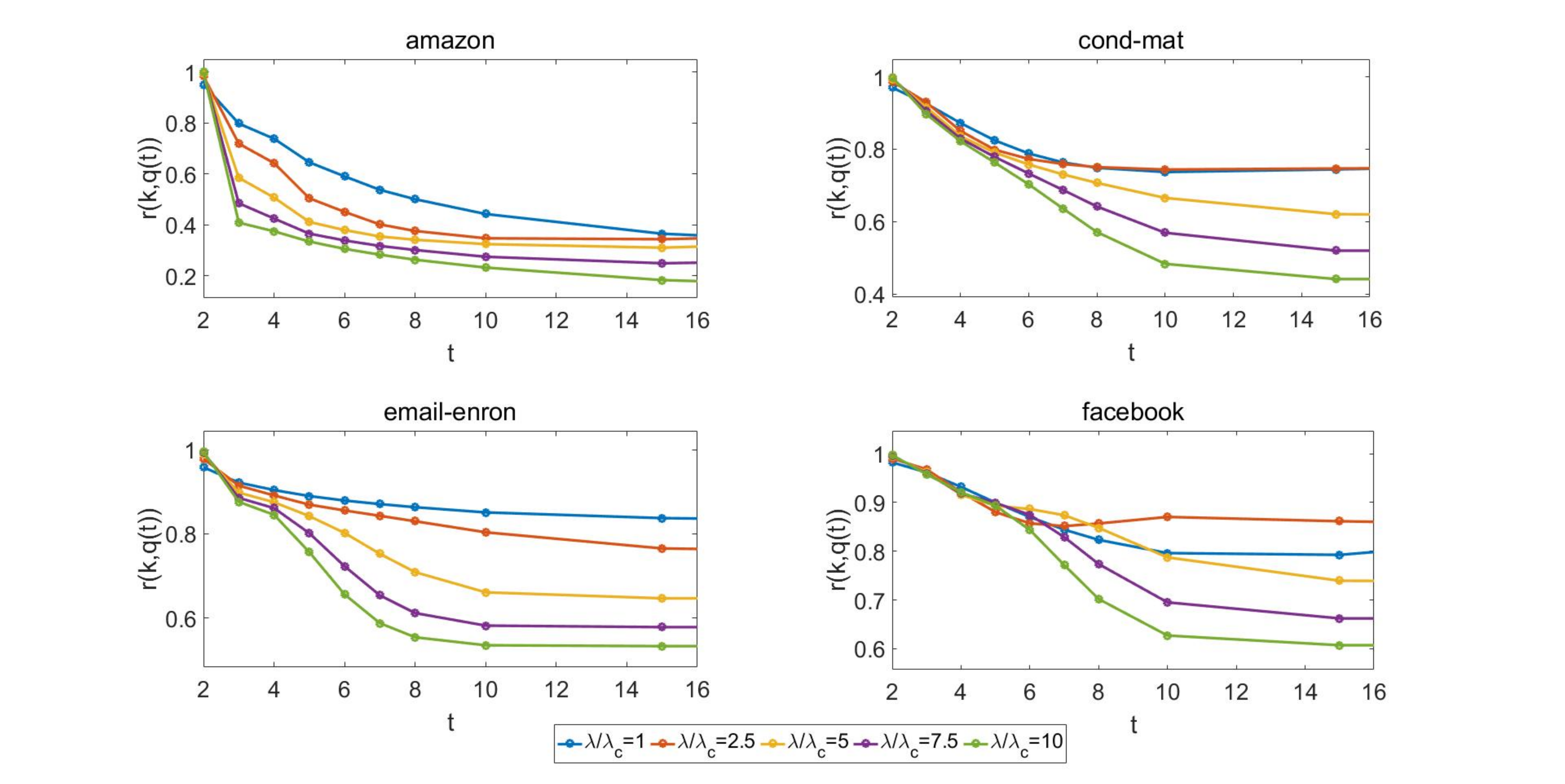}
\caption{Correlation between degree $k$ and $q(t)$ for the four largest networks analyzed here. The correlation tends to decrease with time. }
\label{fig:deg}
\end{figure*}

After two steps of the dynamics, the degree of the nodes is the most relevant property to infer nodes' influence: indeed, we expect $q_i(2)=\beta k_i / N$. The accurate correspondence is lost at longer times, when the local structure around the node of interest plays a non-negligible role, which results in a decreasing correlation between $k$ and $q(t)$ as a function of $t$ (see Fig.~\ref{fig:deg}).
How to exploit network structure to quantify node influence at a time $t>2$ such that the process has not yet reached its final state? 
In other words, which is the most accurate metric to estimate nodes' early-time influence?

One would expect that at early times, it is sufficient to analyze local properties around the focal node. However, there are many structural properties that one could consider, and it is not clear a priori which is the most parsimonious way to accurately gauge node influence. Based on numerical simulations, previous studies~\cite{chen2013identifying,peres2014impact} have pointed out high local clustering as a detrimental property for the influence of a node. Analytically, one can estimate the expected influence of the nodes, which led to the Dynamics-Sensitive Centrality (DSC, see Methods)~\cite{liu2016locating}; however, in principle, its computation requires a detailed knowledge of the parameters of the dynamics.

On the other hand, here, we find that a parsimonious local property that determines node early-time influence is the sum of a node's degree and neighbors' degree: 
\begin{equation}
s_i=k_i + \sum_{j}A_{ij}\,k_j.
\label{social_capital}
\end{equation}
We interpret this property as a centrality metric, which we refer to as \emph{social capital}: according to it, a node is central if it has many connections (i.e., large degree), and if its contacts have many connections as well (i.e., large $\sum_{j}A_{ij}\,k_j$).
Such a social capital function has been used in the context of strategic network formation model~\cite{bardoscia2013social,konig2014nestedness}, where it turns out that a social interaction dynamics where the agents seek to maximize their social capital $s$ leads to highly-centralized network topologies~\cite{bardoscia2013social}.
The metric can be also interpreted as a special case of the Dynamics Sensitive Centrality~\cite{liu2016locating} and of the Neighborhood-based centrality~\cite{liu2016identify}.

For $t>2$, at early times, a temporal regime emerges where the time-dependent influence of the hubs can be accurately quantified by their social capital (see Fig.~\ref{fig:hubs}E-H for an illustration).
The nodes with large social capital tend to achieve high early-time influence, regardless of their local clustering coefficient. 
At late times, the correspondence between node influence and social capital is either lost (see Fig.~\ref{fig:hubs}I,J,L) or, to some extent, maintained (see Fig.~\ref{fig:hubs}K). This set of observations suggests that social capital can be an effective method for the identification of fast influencers, a hypothesis that we validate below.

\section{Fast and late-time influencer identification: benchmarking centrality metrics}
\label{sec:benchmarking}

The goal of this Section is to benchmark the centrality metrics introduced in Section~\ref{sec:centrality} with respect to their ability to single out the fast influencers (Section~\ref{sec:fast}) and the late-time influencers (Section~\ref{sec:late_time}).
Finally, we discuss the complementary nature of local and global centrality metrics (Section~\ref{sec:complementarity}).

\subsection{Identification of fast influencers}
\label{sec:fast}

\begin{figure*}[t]
\centering 
\includegraphics[scale=0.23]{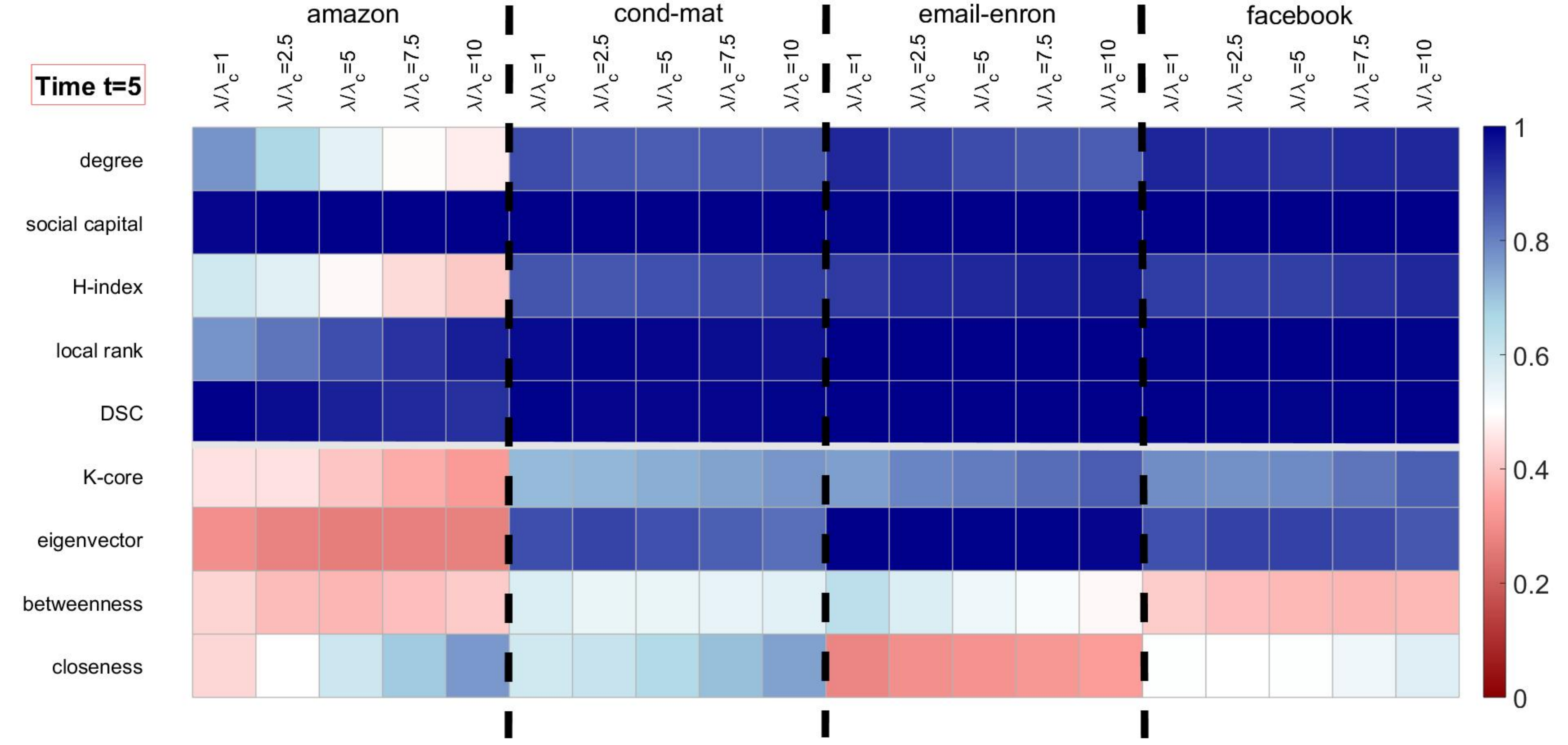}
\caption{Metrics' ability to identify the fast influencers, as measured by the linear correlation
between node influence $q(5)$ after $5$ steps of the dynamics and node score as determined by the nine centrality metrics considered here. The color scale 
ranges from dark red (zero correlation) to dark blue (correlation achieved by the best-performing metrics);
the white color denotes half of the correlation achieved
by the best-performing metric. For all the analyzed datasets and parameter values, the correlation achieved 
by Social Capital is always the largest one or close to 
the largest one.
}
\label{fig:short_term}
\end{figure*}

\begin{figure*}[t]
\centering
\includegraphics[scale=0.23]{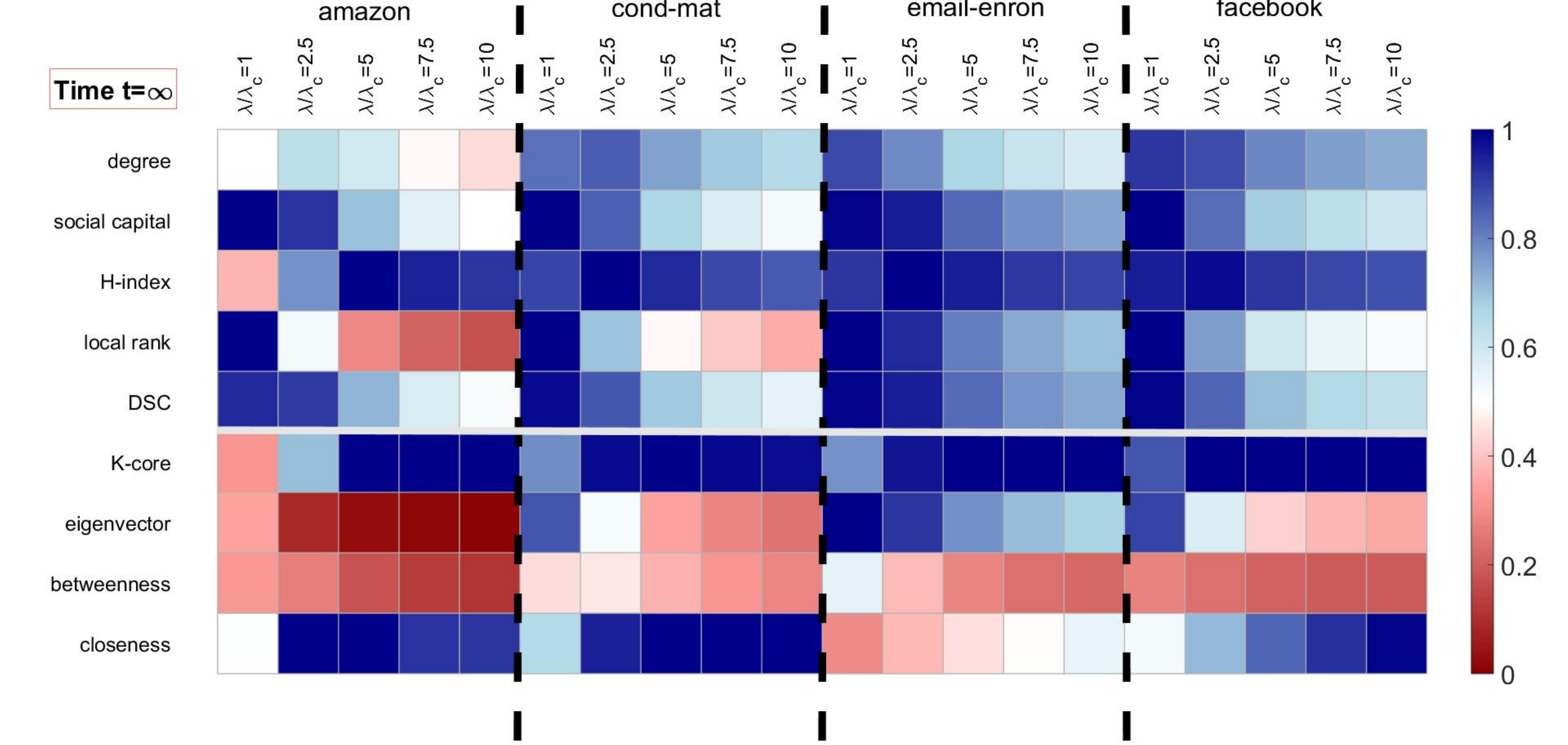}
\caption{Metrics' ability to identify the late-time influencers, as measured by the linear correlation between node late-time influence $q(\infty)$ and node score as determined by the nine centrality metrics considered here. The color scale ranges from dark red (zero correlation) to dark blue (correlation achieved by the best-performing metrics); the white color denotes half of the correlation achieved by the best-performing metric. $k$-core turns out to be competitive for a broad range of parameter values, yet it fails at the critical point $\lambda=\lambda_c$ in Amazon. Despite its local nature, social capital is the only metric whose correlation did not fall below half of the best-performing correlation for any of the considered parameter values.}
\label{fig:late_time}
\end{figure*}

\begin{figure*}[t]
\centering
\includegraphics[scale=0.23]{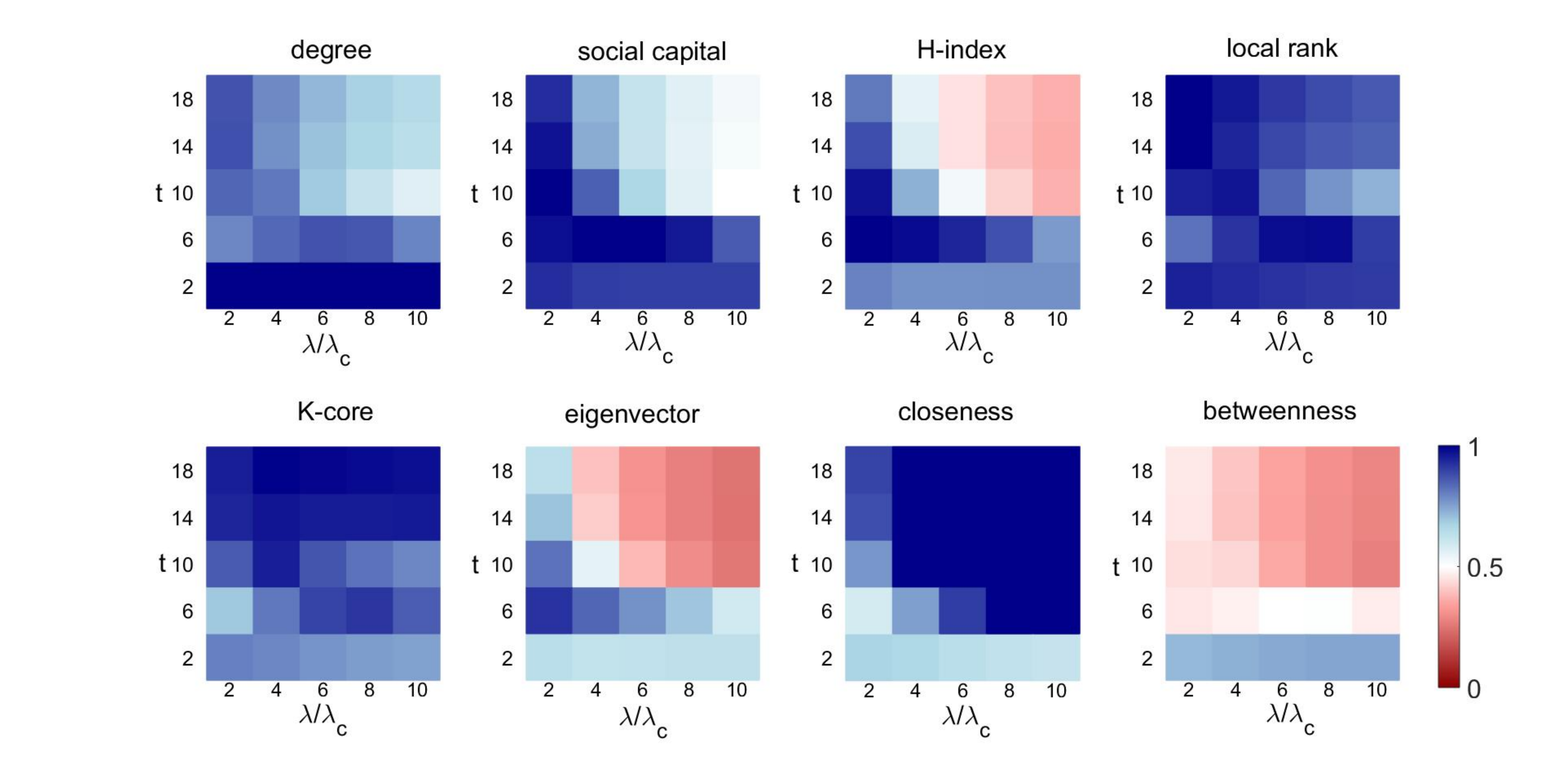}
\caption{Metrics' performance in the \emph{Cond-mat} network as measured by their correlation with node influence at different times $t$ and for different $\lambda$ values. Local metrics like Social Capital and $H$-index tend to perform better in the lower-left corner of the heatmap (early time, processes close to the critical point), whereas global metrics like the $k$-core and closeness centrality tend to perform better in the upper-right corner of the heatmap (late time, processes sufficiently far from the critical point). A similar qualitative picture holds for other datasets (see~\ref{appendix:complementarity}).
}
\label{fig:lambdat}
\end{figure*}

In the following, we investigate the performance of different metrics for the vital nodes identification~\cite{lu2016vital} at early times. We consider nine centrality metrics: degree, social capital (defined above), $H$-index, LocalRank, Dynamics Sensitive Centrality (DSC), $k$-core, eigenvector centrality, betweenness centrality, closeness centrality. We refer to the Methods section for all the definitions. To evaluate the metrics, we measure both the precision $P_{0.5\%}(s,q(t))$ of the ranking by the score $s$ that they produce in identifying the top-$0.5\%$ nodes by influence $q(t)$, and the Pearson's linear correlation coefficient $r(s, q(t))$ between their score $s$ and node influence. In the following, we always normalize the observed values of correlations and precision by the correlation and precision values, respectively, achieved by the best-performing metric.
In this Section, we focus on the identification of fast influencers, whereas Section~\ref{sec:late_time} focuses on the identification of late-time influencers.

Fig.~\ref{fig:short_term} compares the correlation $r(\cdot,q(5))$ between node influence at time $5$ and node score achieved by the various centrality metrics, for the four largest analyzed networks. Social capital turns out to be always either the best performing metric or relatively close to the best. The DSC and the LocalRank centrality perform similarly well, yet they are marginally outperformed by the Social Capital centrality in \emph{Amazon}.
By contrast, the correlation values achieved by the other eight metrics are, at least once, smaller than half of the correlation achieved by the best-performing metric. Overall, local metrics substantially outperform global metrics -- the only exception is the nearly-optimal performance of the eigenvector centrality in \emph{Email-Enron}.
Qualitatively similar results hold for the precision metric (see~\ref{appendix:precision}).

%This can be observed by considering the largest network analyzed in this paper, Amazon. In Amazon, the correlation of all metrics but social capital is below half of the correlation of the best-performing metric for at least one parameter value: degree for $\lambda>=5\,\lambda_c$; $H$-index for $\lambda<=\lambda_c$; LocalRank for $\lambda>=7.5\,\lambda_c$; DSC for $\lambda>=7.5\,\lambda_c$; $k$-core for all the studied values of $\lambda$; eigenvector for $\lambda<=\lambda_c$,
%betweenness for all the studied values of $\lambda$; closeness for $\lambda=\lambda_c$. Interestingly, closeness is the best-performing metric for $\lambda>=7.5\,\lambda_c$, which supports the idea that distance-based centrality metrics can only be competitive sufficiently above the critical point~\cite{iannelli2018network}.

While Fig.~\ref{fig:short_term} looked at a specific point in time ($t=5$) for four empirical networks, two extensions of the analysis are essential. First, it is important to investigate the stability of the relative metrics' performance at early times. This aspect is the main topic of Section~\ref{sec:complementarity}.
Second, one can consider more datasets. We analyzed ten more networks, ending up with a total of $14$ analyzed networks (the networks' details are reported in Table~\ref{tab}). 
We aggregated the performance of each metric over the $14$ networks according to the following procedure. For each network $\mathcal{D}$ and metric $m$, we normalized the metric's correlation with node influence $q(t)$, $r(m,t;\mathcal{D})$, by the correlation $r_{best}(t;\mathcal{D})$ achieved by the best-performing metric. Then, we averaged the metrics relative performance as $|\mathcal{D}|^{-1}\sum_{\mathcal{D}} r(m,t;\mathcal{D})/r_{best}(t;\mathcal{D})$, where $|\mathcal{D}|$ is the number of analyzed datasets.
The results of this analysis (shown in~\ref{appendix:aggregate}) are in qualitative agreement with our results in Fig.~\ref{fig:short_term}: at early times ($t=5$), social capital, LocalRank and the DSC are, on average, the best performing metrics.

\subsection{Identification of late-time influencers through centrality metrics}
\label{sec:late_time}

At late times, the identification of the best-performing metric is strongly dataset-dependent and parameter-dependent.
At the critical point, $\lambda=\lambda_c$, Social Capital, LocalRank the Dynamics Sensitive Centrality tend to be the best-performing metrics according to both precision (see~\ref{appendix:precision}) and correlation (see Fig.~\ref{fig:late_time}). 
On the other hand, if $\lambda$ is sufficiently large (e.g., $\lambda=5\,\lambda_c$), all metrics' precision sharply decreases after a sufficiently long time due to the fact that most nodes become infected, regardless of the seed node (see~\ref{appendix:precision}). Even when this happens, it remains possible to discriminate the metrics' performance based on the correlation metric (see Fig.~\ref{fig:late_time}).
The $k$-core's performance is maximal or nearly maximal for $\lambda\geq 5\,\lambda_c$; for both \emph{Cond-mat} and \emph{Facebook}, the closeness centrality performs optimally or nearly optimally for $\lambda\geq 5$, which supports the competitiveness of distance-based centrality metrics for processes that are sufficiently far from the critical point~\cite{iannelli2018influencers}.

Again, to obtain a broader understanding of the metrics' performance, one needs to consider more datasets.
As a result of an analogous analysis to the one described in the last paragraph of Section~\ref{sec:fast}, by analyzing $14$ networks in total, we find that
at $t\to\infty$, $k$-core and $H$-index are the most competitive metrics according to the average correlation (see~\ref{appendix:aggregate}). The same does not hold for precision, where the metrics' performance is impaired by the fact that the late-time influence of most nodes approaches one. Again, closeness can only be competitive for sufficiently large values of $\lambda$.

%\begin{figure*}
%\includegraphics[scale=0.3]{figs/aggegrate_metric_14_networks_score.pdf}
%\caption{mean relative metric precision score of total 14 networks}

%\includegraphics[scale=0.29]{figs/aggegrate_metric_14_networks_rank.pdf}
%\caption{mean rank by metric precision score of total 14 networks}
%\end{figure*}{top}
%\begin{figure*}
%\centering
%\includegraphics[scale=0.3]{figs/correlati%on_aggegrate_relative_score.pdf}
%\caption{mean relative metric correlation score of total 14 networks.}

%\includegraphics[scale=0.3]{figs/correlation_aggegrate_rank.pdf}
%\caption{mean rank by metric correlation score of total 14 networks.}

%\end{figure*}

\subsection{The complementarity of local and global metrics}
\label{sec:complementarity}

In the previous results, we quantified the nodes' early-time influence through their influence at step $t=5$, $q(5)$. Yet, it is important to investigate the performance of the centrality metrics in the full $(\lambda,t)$ plane. The results (shown in Figs.~\ref{fig:lambdat}, \ref{fig:lambdat2}-\ref{fig:lambdat3} in~\ref{appendix:complementarity}) indicate the complementary nature of (well-performing) local and global metrics in quantifying node influence. While local metrics (degree, social capital, $H$-index, and LocalRank) achieve optimal or nearly optimal performance for small $t$ and small $\lambda$, global metrics ($k$-core and closeness centrality) emerge as optimal or nearly optimal metrics in the parameter region that corresponds to large $t$ and large $\lambda$.

This scenario changes qualitatively for the precision metric. However, for the precision metric, the evaluation in the full parameter space is impaired by the fact that for some datasets, as node influence becomes close to one for most metrics, the precision of all the considered metrics is relatively low. This is evident in Figs.~\ref{fig:prec1}-\ref{fig:prec3} in~\ref{appendix:complementarity}, where we use a black color to mark the parameter values where all the metrics achieve a precision value smaller than 10\%. For the remaining parameter values (i.e., those where at least one metric is able to achieve a precision larger than 10\%), local metrics (social capital, $H$-index, LocalRank) typically outperform global metrics ($k$-core, eigenvector, and closeness centrality). 

\section{Conclusions}
\label{sec:conclusions}

We have investigated the impact of time on the long-standing influence maximization problem, and performed an extensive benchmarking of centrality metrics based on their performance in the identification of both fast and late-time influencer nodes. Our results revealed that the early-time and late-time influence of a node can differ substantially. Besides, a simple local metric -- which we referred to as social capital -- allows us to identify the fast influencers substantially better than global metrics, by using less information compared to existing local metrics (such as LocalRank~\cite{chen2012identifying} and Dynamics-Sensitive Centrality~\cite{liu2016locating}).

In the debate on the relative usefulness of local and global centralities~\cite{lu2016vital,hu2018local}, our results indicate that local metrics generally outperform global metrics at early times and for processes with a small spreading rate, whereas global metrics may still outperform local metrics at late-times or for processes with a large spreading rate.
Beyond epidemiological models, it remains open to investigate the fast influencer identification problem in complex contagion~\cite{kempe2003maximizing} and rumor spreading models~\cite{borge2012absence,borge2013emergence}.
Nevertheless, the main message of our paper still holds: when assessing the performance of metrics for the influencer identification, nodes' early- and late-time influence can be only weakly correlated, which makes it critical to assess how the metrics' relative performance depends on the considered timescale. 

To conclude, our work investigated the essential role of time in the vital nodes identification problem.
Our findings point out that given a complex network, the optimal late-time influencers might be suboptimal if we are interested maximizing the reach of a spreading process within a limited amount of time.
An important future direction is the extension of these results to temporal networks~\cite{holme2016temporal}, where metrics for node influence can be biased by temporal effects~\cite{mariani2015ranking,liao2017ranking}, and memory effects have a fundamental impact on diffusion processes~\cite{scholtes2014causality,xu2016representing} and epidemic spreading~\cite{holme2016temporal}. 

\clearpage

\appendix

\section{Results for the precision metric}
\label{appendix:precision}

\begin{figure*}[h]
\includegraphics[scale=0.27]{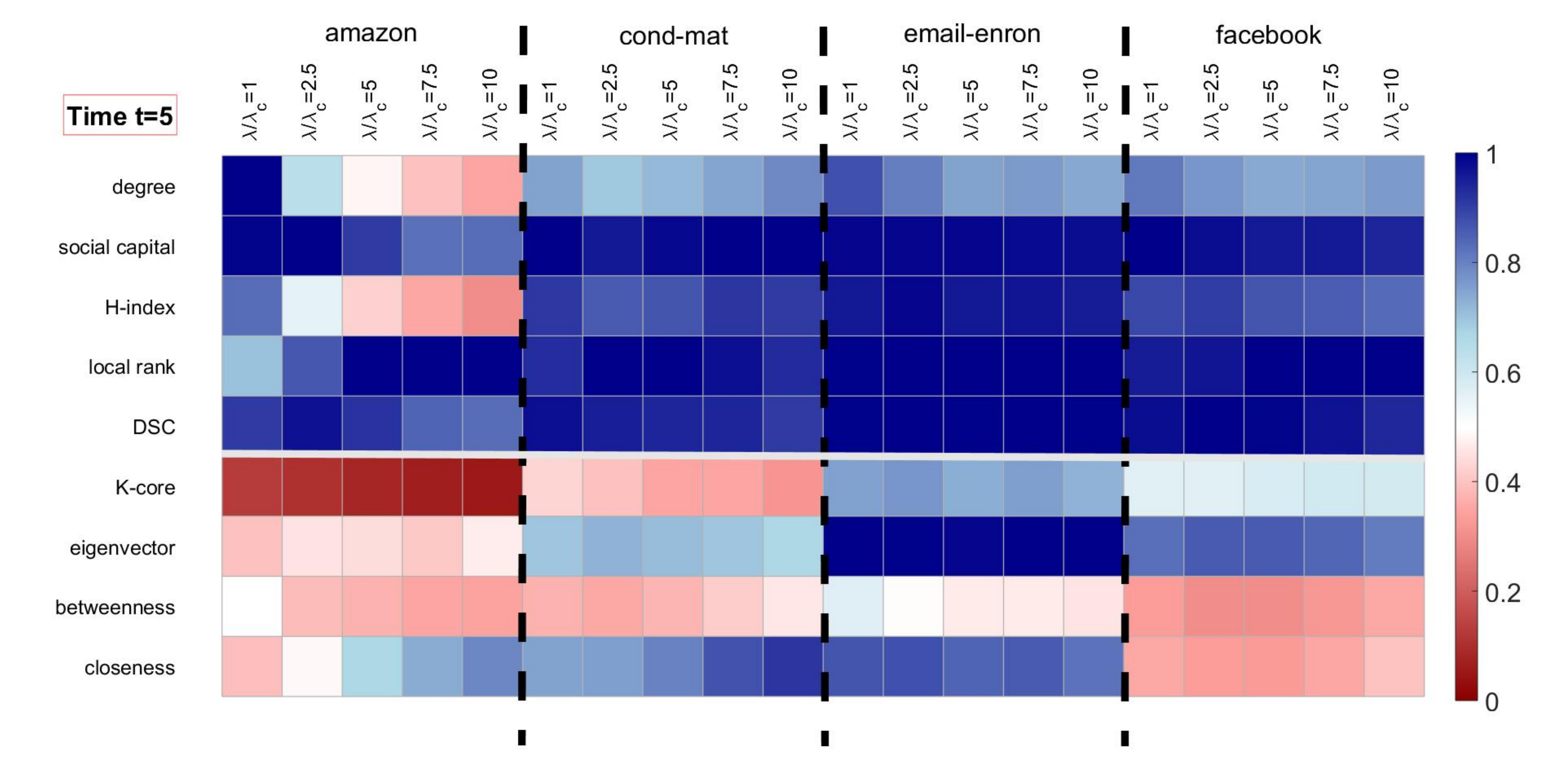}
\caption{Metrics' ability to identify 
the fast influentials, as measured
by the metrics' precision in identifying the top-$0.5\%$ nodes by short-time influence $q(5)$. The color scale ranges from dark red (zero precision) to dark blue 
(precision achieved by the best-performing metrics); the white 
color denotes half of the precision achieved by the best-performing metric. 
All the metrics but Social Capital, LocalRank, and the DSC have at least 
one white or red box, which indicates that Social Capital is a parsimonious
and effective method to identify the 
fast influentials.}
\end{figure*}

\begin{figure*}[h]
\includegraphics[scale=0.269]{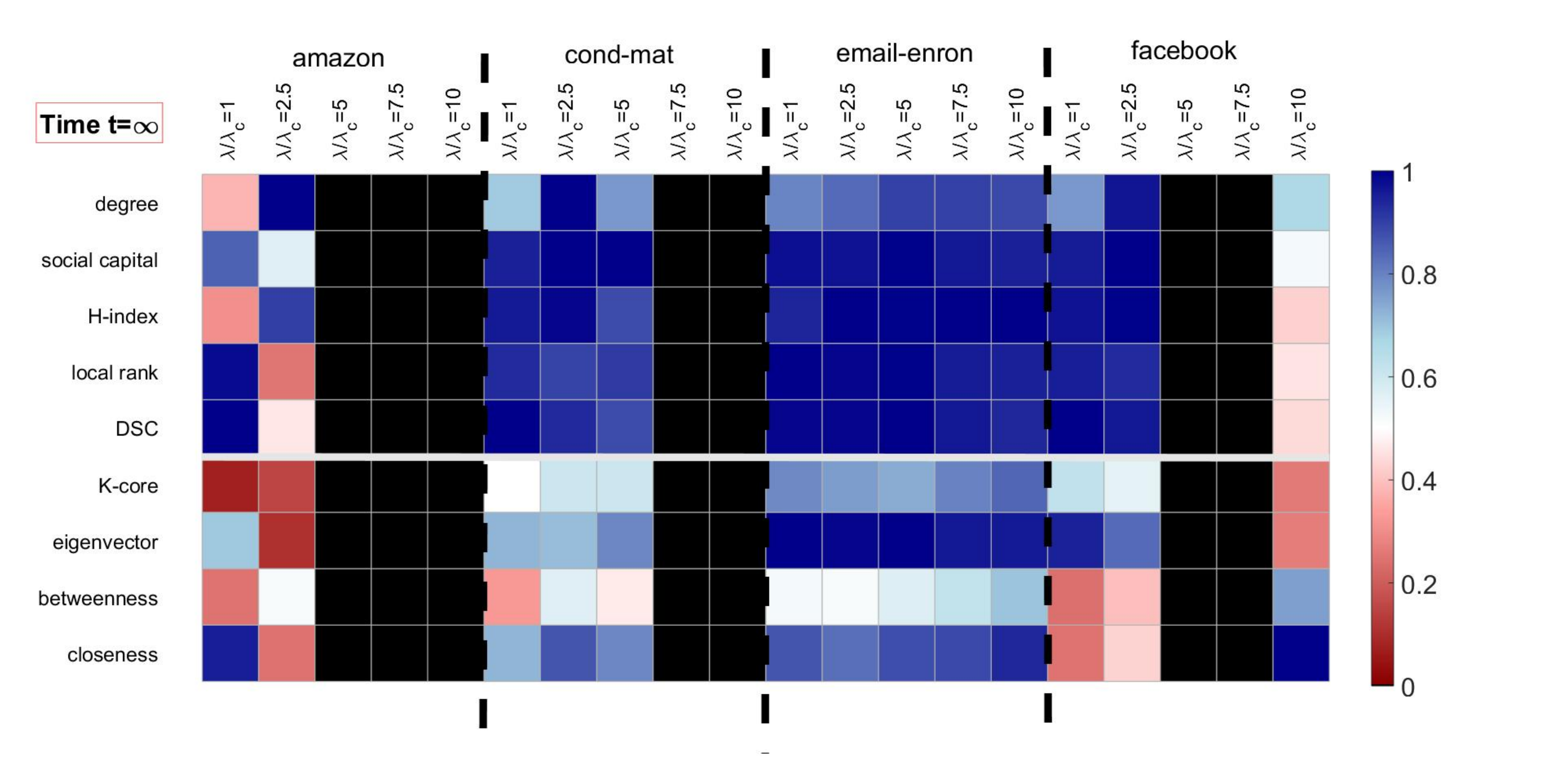}
\caption{Metrics' ability to identify the late-time influentials, as measured by the metrics' precision in identifying the top-$0.5\%$ nodes by late-time influence $q(\infty)$. The color scale ranges from dark red (zero precision) to dark blue (precision achieved by the best-performing metrics); the white color denotes half of the precision achieved by the best-performing metric. We marked in black those parameter values for which no metric achieves a precision larger than $10\%$. $k$-core turns out to be competitive for a broad range of parameter values, yet it fails at the critical point $\lambda=\lambda_c$ in Amazon. Despite its local nature, social capital is the only metric whose correlation did not fall below half of the best-performing correlation, for the considered parameter values. }
\end{figure*}

\clearpage

\section{Metrics' aggregate performance over 14 networks}
\label{appendix:aggregate}

\begin{figure*}[h]
\centering
\includegraphics[scale=0.28]{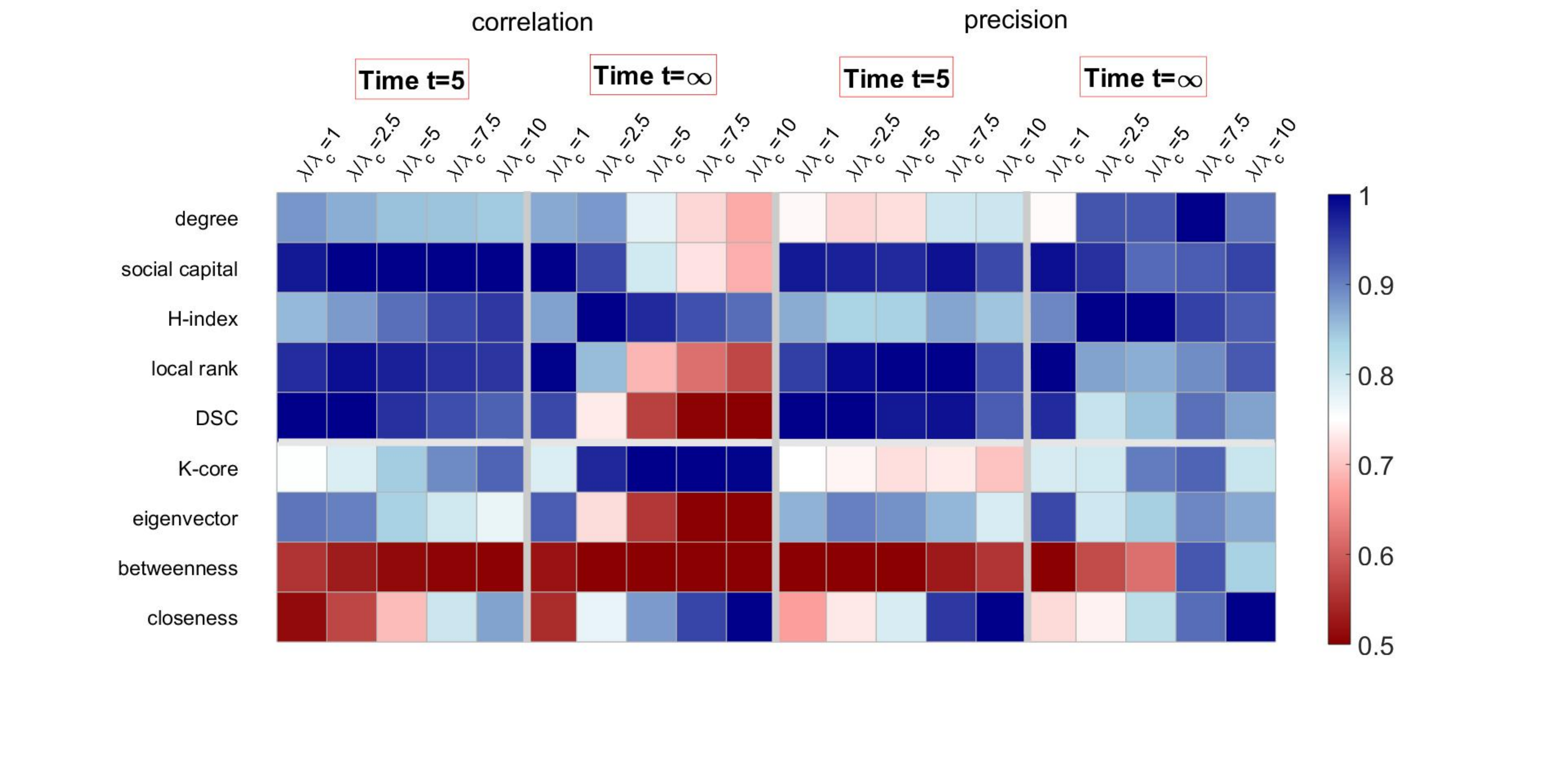}
\caption{Average relative metrics' performance over all the datasets
in Table~\ref{tab}. At early times ($t=5$), social capital, LocalRank and the DSC emerge as the best performing metrics. At late times ($t=\infty$), $k$-core and $H$-index are the most competitive metrics according to correlation, whereas degree, Social Capital and $H$-index perform similarly well according to precision. The closeness centrality is only competitive for $\lambda=10\,\lambda_c$.}
\label{fig:aggregate}
\end{figure*}

We follow the procedure described in the main text: For each dataset $\mathcal{D}$ and each metric $m$, we normalize the metric's correlation (with node influence) $r(m;\mathcal{D})$ by the correlation $r_{best}(\mathcal{D})$ achieved by the best-performing metric for that dataset. For each pair $(\lambda,t)$, we obtain the average relative performance of the metric as $|\mathcal{D}|^{-1}\sum_{d\in\mathcal{D}}r(m;\mathcal{D})/r_{best}(\mathcal{D})$.
We follow an analogous procedure for the metrics' precision.
The results are shown and commented in Fig.~\ref{fig:aggregate}.

\clearpage

\section{The complementarity of local and global metrics: Additional results}
\label{appendix:complementarity}

\begin{figure*}[h]
\includegraphics[scale=0.23]{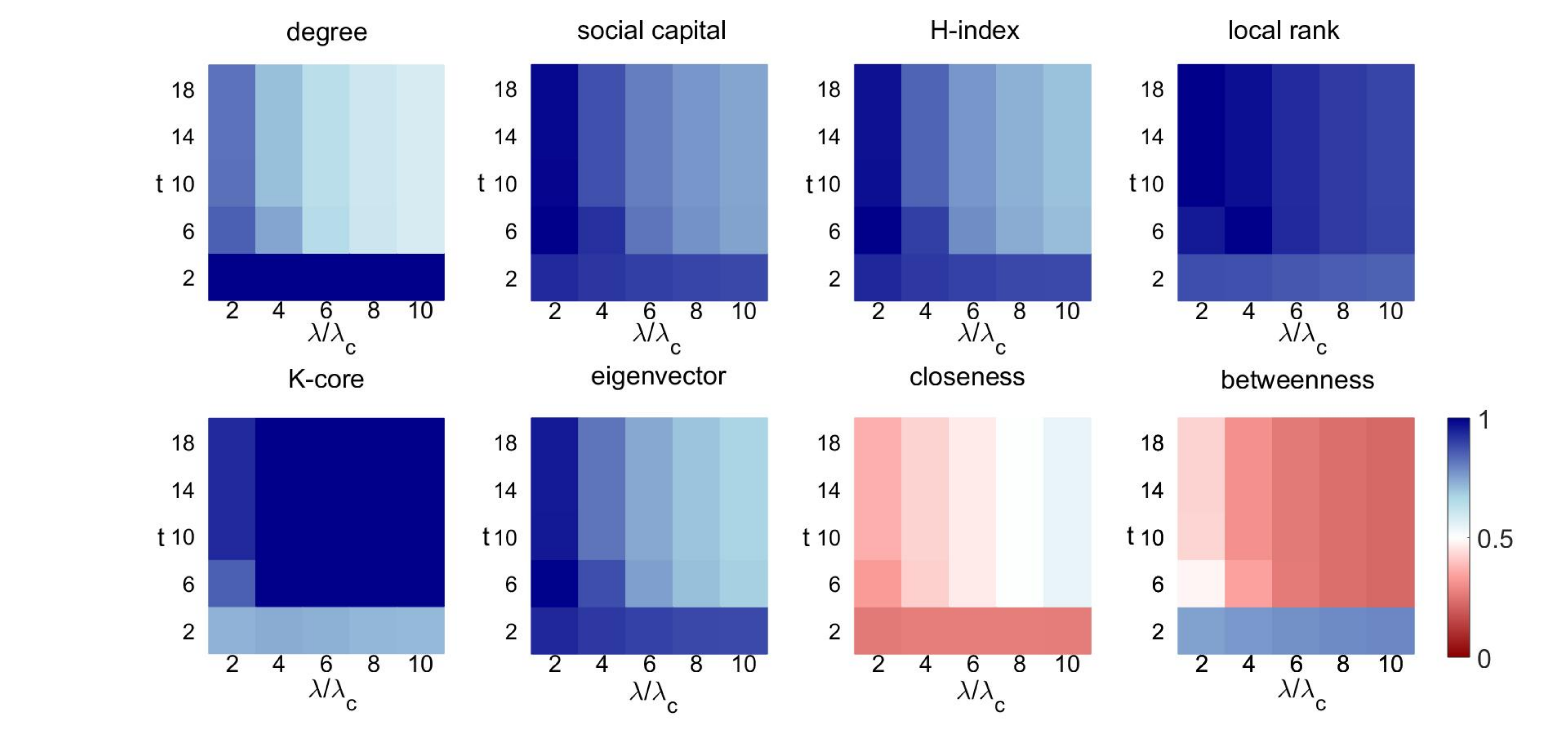}
\caption{Metrics' performance in the \emph{Email-Enron} network as measured by their correlation with node influence at different times $t$ and for different $\lambda$ values. Local metrics like Social Capital and $H$-index tend to perform better in the lower-left corner of the heatmap (early time, processes close to the critical point), whereas global metrics like the $k$-core and closeness centrality tend to perform better in the upper-right corner of the heatmap (late time, processes sufficiently far from the critical point).}
\label{fig:lambdat2}
\end{figure*}

\begin{figure*}[h]
\includegraphics[scale=0.23]{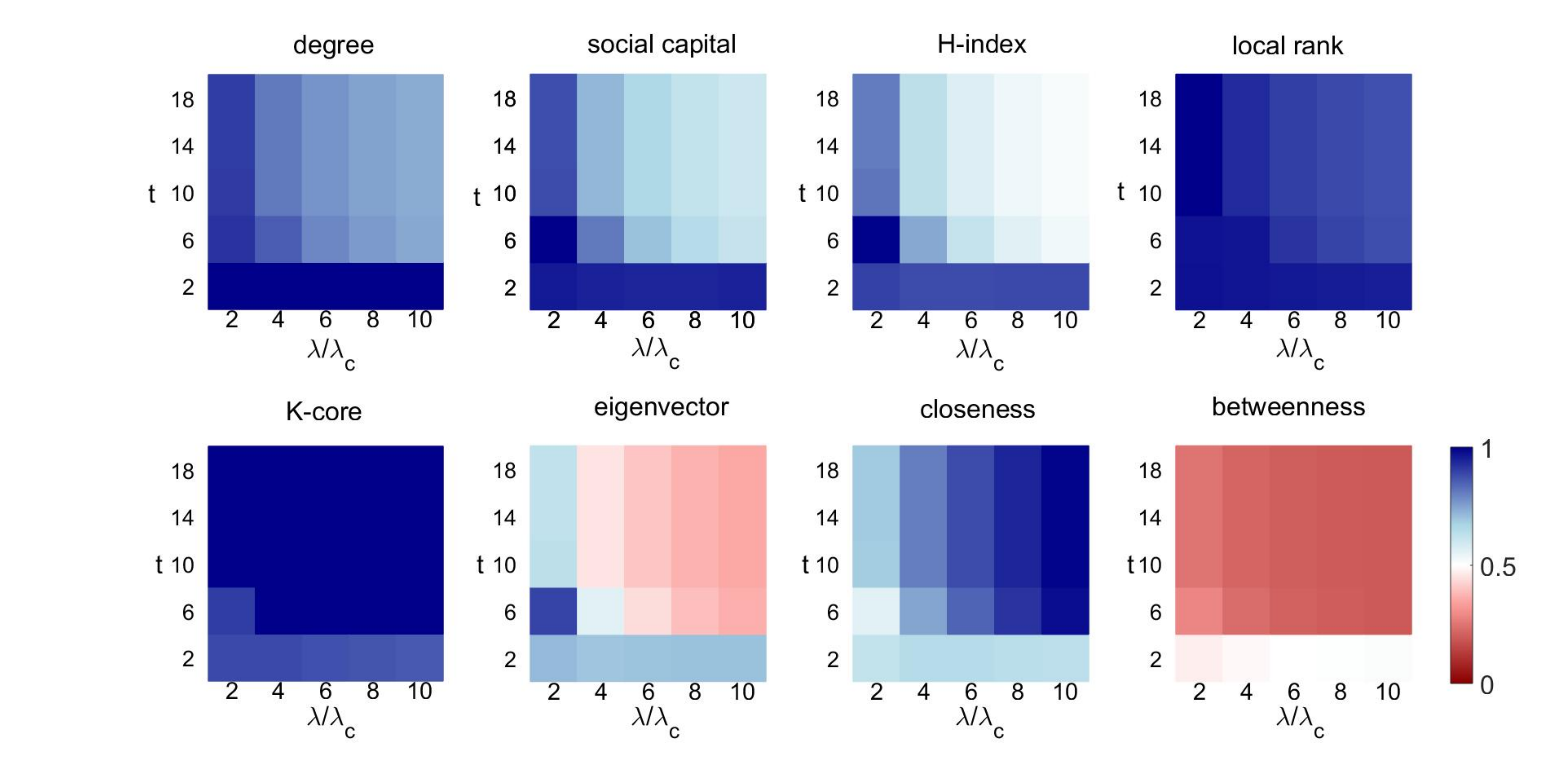}
\caption{Metrics' performance in the \emph{Facebook} network as measured by their correlation with node influence at different times $t$ and for different $\lambda$ values. Local metrics like Social Capital and $H$-index tend to perform better in the lower-left corner of the heatmap (early time, processes close to the critical point), whereas global metrics like the $k$-core centrality tend to perform better in the upper-right corner of the heatmap 
(late time, processes sufficiently far from the critical point).}
\label{fig:lambdat3}
\end{figure*}

\begin{figure*}[h]
\includegraphics[scale=0.26]{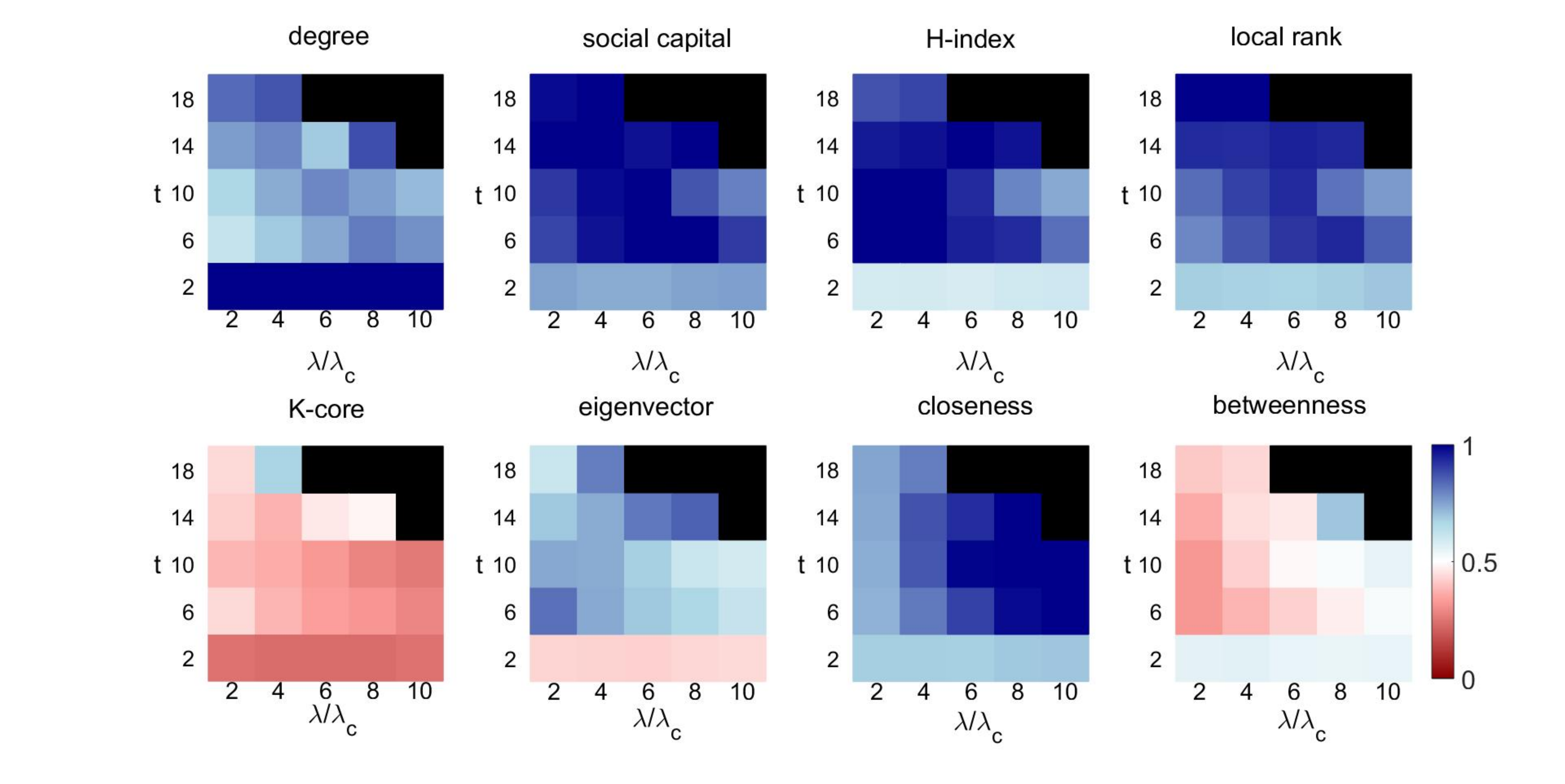}
\caption{Metrics' performance in the \emph{Cond-mat} network as measured by their precision in identifying the top-$0.5\%$ nodes by influence at different times $t$ and for different $\lambda$ values. The block color marks boxes that correspond to $(\lambda,t)$ values where all the metrics' precision is below $0.1$.}
\label{fig:prec1}
\end{figure*}

\begin{figure*}[h]
\includegraphics[scale=0.26]{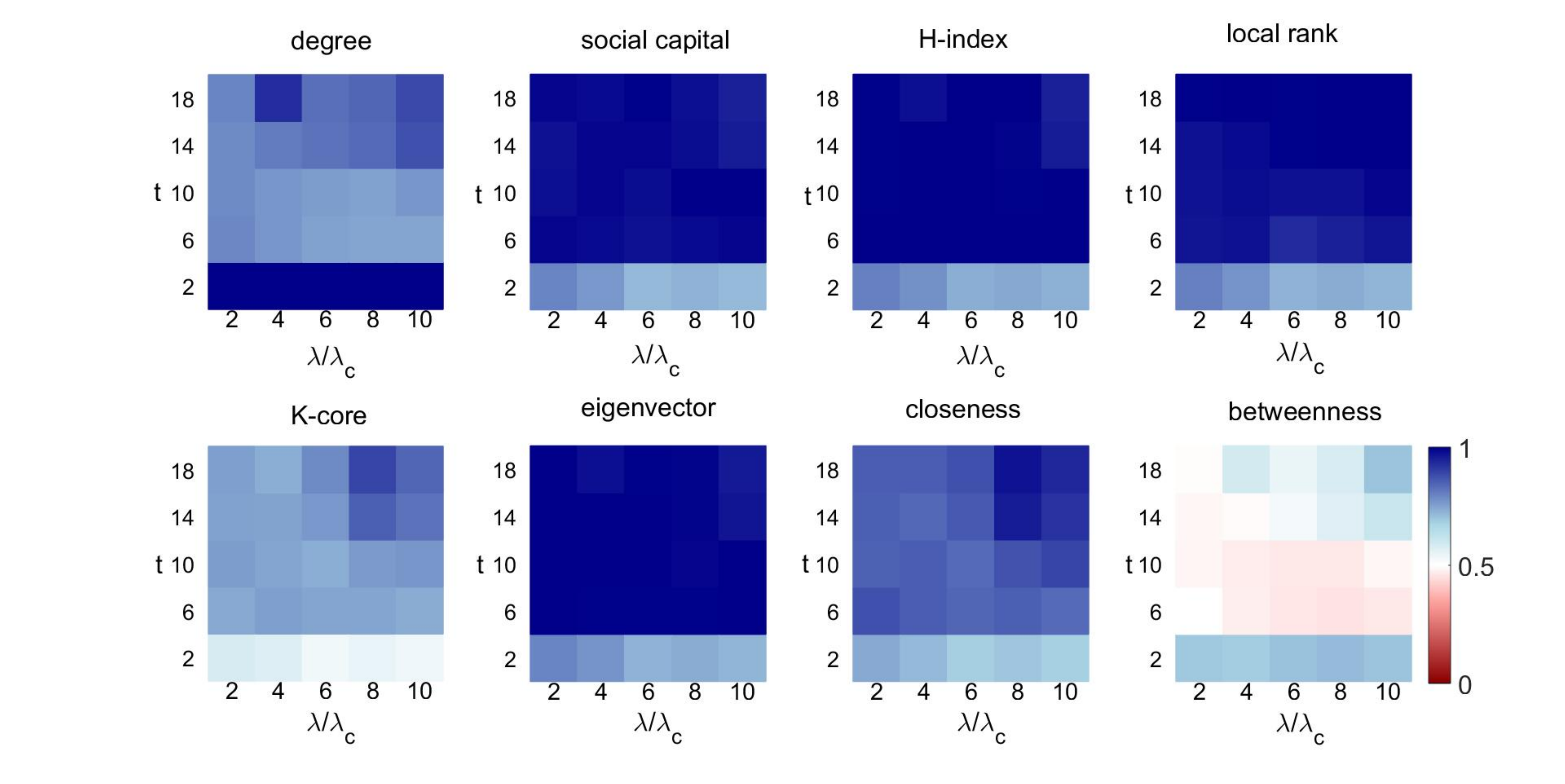}
\caption{Metrics' performance in the \emph{Email-Enron} network as measured by their precision in identifying the top-$0.5\%$ nodes by influence at different times $t$ and for different $\lambda$ values. The block color marks boxes that correspond to $(\lambda,t)$ values where all the metrics' precision is below $0.1$.}
\label{fig:prec2}
\end{figure*}

\begin{figure*}[h]
\includegraphics[scale=0.26]{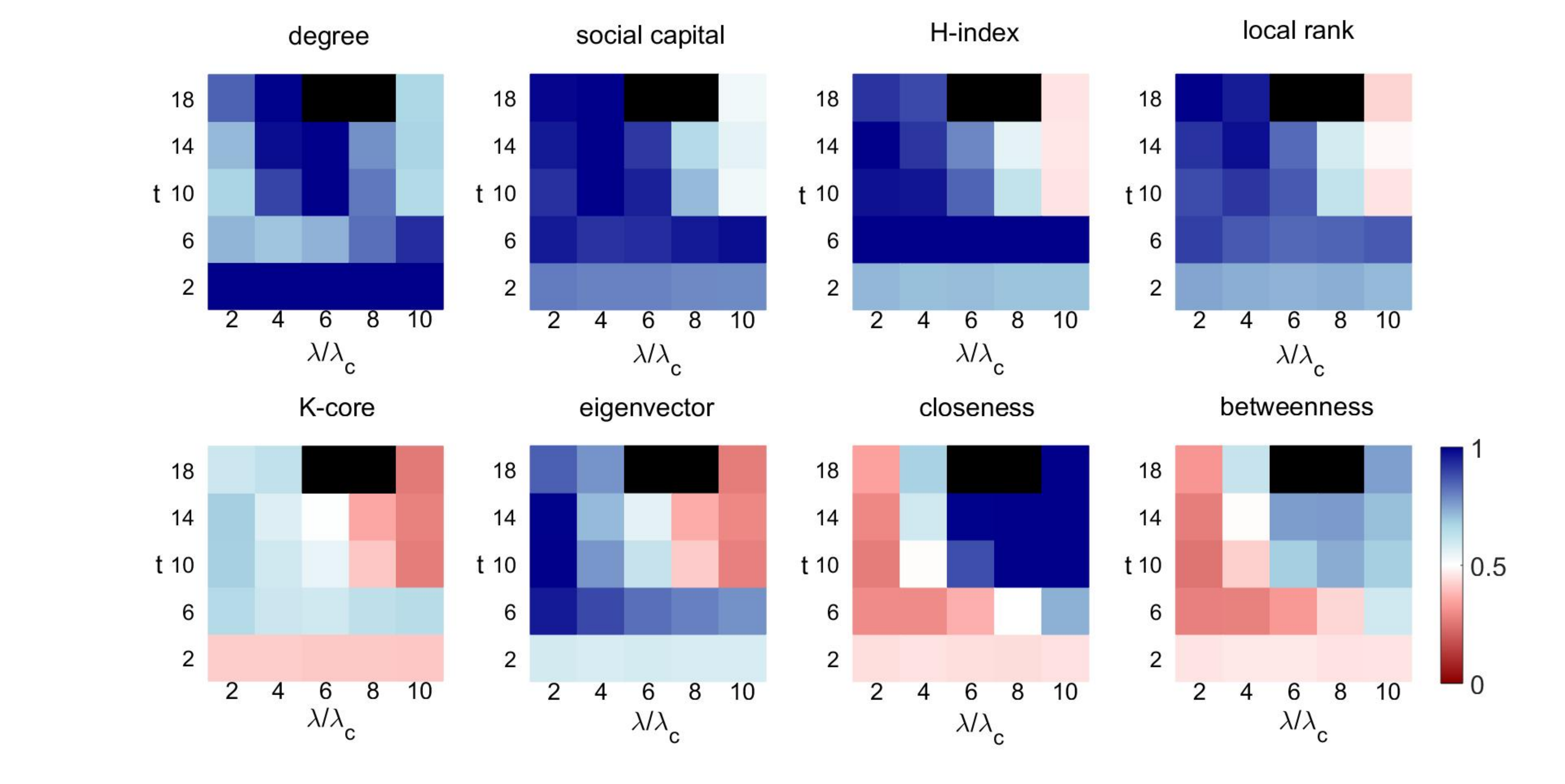}
\caption{Metrics' performance in the \emph{Facebook} network as measured by their precision in identifying the top-$0.5\%$ nodes by influence at different times $t$ and for different $\lambda$ values. The block color marks boxes that correspond to $(\lambda,t)$ values where all the metrics' precision is below $0.1$.}
\label{fig:prec3}
\end{figure*}

\clearpage 

\section*{Acknowledgments}
We thank Flavio Iannelli for many enlightening discussions on the topic, and his feedback on an early version of the manuscript. This work has been supported by the National Natural Science Foundation of China (Grants Nos. 61673150, 11622538), the Science Strength Promotion Program of the UESTC, and the 
Zhejiang Provincial Natural Science Foundation of China (Grant no. LR16A050001). MSM acknowledges the University of Z\"urich for support through the URPP Social Networks.

\end{document}